\documentclass[iop]{emulateapj}
\usepackage{color}

\slugcomment{Draft version}

\shorttitle{Distant CO-emitting galaxy beyond VV114}
\shortauthors{Tamura et al.}

\begin{document}


\title{Serendipitous ALMA detection of 
a distant CO-emitting galaxy \\with 
a buried active galactic nucleus beyond 
the nearby merging galaxies VV114}


\author{
  Yoichi Tamura\altaffilmark{1}, 
  Toshiki Saito\altaffilmark{2,3}, 
  Takeshi~G.\ Tsuru\altaffilmark{4}, 
  Hiroyuki Uchida\altaffilmark{4}, 
  Daisuke Iono\altaffilmark{2}, 
  Min~S. Yun\altaffilmark{5}, 
  Daniel Espada\altaffilmark{6}, 
  and 
  Ryohei Kawabe\altaffilmark{6,2}
}


\altaffiltext{1}{Institute of Astronomy, The University of Tokyo, Mitaka, Tokyo 181-0015, Japan}
\altaffiltext{2}{National Astronomical Observatory of Japan, Mitaka, Tokyo 181-8588, Japan}
\altaffiltext{3}{Department of Astronomy, The University of Tokyo, Hongo, Bunkyo-ku, Tokyo 113-0033, Japan}
\altaffiltext{4}{Division of Physics and Astronomy, Kyoto University, Kitashirakawa-Oiwake-Cho, Sakyo, Kyoto 606-8502, Japan}
\altaffiltext{5}{Department of Astronomy, University of Massachusetts, 710 North Pleasant Street, Amherst, MA 01003, USA}
\altaffiltext{6}{Joint ALMA Observatory, Alonso de Cordova 3107, Vitacura, Santiago 763 0355, Chile}


\begin{abstract}
We report the detection of a distant star-forming galaxy, ALMA J010748.3$-$173028, which is identified by a 13$\sigma$ emission line at 99.75 GHz ($S\Delta v = 3.1$ Jy km s$^{-1}$), behind 
the nearby merging galaxies VV114 using the Atacama Large Millimeter/submillimeter Array (ALMA) Band 3.  We also find an 880-$\micron$ counterpart with ALMA Band~7 ($S_\mathrm{880\mu m} = 11.2$~mJy).  A careful comparison of the intensities of the line and the continuum suggests that the line is a redshifted $^{12}$CO transition.  A photometric redshift analysis using the infrared to radio data favors a CO redshift of $z = 2.467$, although $z = 3.622$ is acceptable.
We also find a hard X-ray counterpart, suggesting the presence of a luminous ($L_\mathrm{X} \sim 10^{44}$~erg~s$^{-1}$) active galactic nucleus obscured by a large hydrogen column ($N_\mathrm{H} \sim 2 \times 10^{23}$ cm$^{-2}$ if $z = 2.47$).
A cosmological simulation shows that the chance detection rate of a CO-emitting galaxy at $z > 1$ with $\ge 1$ Jy km s$^{-1}$ is $\sim 10^{-3}$ per single ALMA field of view and 7.5-GHz bandwidth at 99.75~GHz.  This demonstrates that ALMA has sufficient sensitivity to find an emission-line galaxy such as ALMA~J010748.3$-$173028 even by chance, although the likelihood of stumbling across such a source is not high. 
\end{abstract}




\section{Introduction}

In recent years, the advent of new millimeter (mm) and submillimeter (submm) facilities, such as the Atacama Large Millimeter/submillimeter Array (ALMA), with unprecedented sensitivity and frequency coverage, has improved the detectability of cool gas in high-redshift star-forming galaxies \citep[e.g.,][]{Wagg12, Vieira13, Wang13}.  \citet{Swinbank12} reported two chance detections of submm galaxies \citep[SMGs,][for a review]{Blain02} in \textsc{[C~ii]} 158-$\micron$ 
line emission during 870-$\micron$ continuum follow-up observations of 126 SMGs using ALMA, which places the first constraint on the \textsc{[C~ii]} luminosity function at $z = 4.4$.  Furthermore, \citet{Hatsukade13} made the deepest unlensed number counts of SMGs using sources incidentally detected in the same fields of view (FoVs) toward near-infrared-selected star-forming galaxies at $z = 1.4$.  Such ``incidental'' searches for high-$z$ star-forming galaxies, especially in emission lines, offer a unique opportunity to investigate the line luminosity functions or cool gas mass functions, which are important for constraining galaxy formation models.

Here we report the serendipitous detection of a dusty starburst galaxy, ALMA~J010748.3$-$173028 (hereafter ALMA-J0107), with a significant emission line at 99.75 GHz, which is likely a redshifted $^{12}$CO line.  At this position, we find a hard X-ray source, which strongly suggests the presence of a buried active galactic nucleus (AGN).

We assume a cosmology with $\Omega_\mathrm{m} = 0.3$, $\Omega_{\Lambda} = 0.7$, and $H_0 = 70$ km s$^{-1}$ Mpc$^{-1}$ ($h = 0.7$).



\begin{figure*}[htbp]
\begin{center}
\includegraphics[angle=0,width = 0.98\textwidth,keepaspectratio,clip]{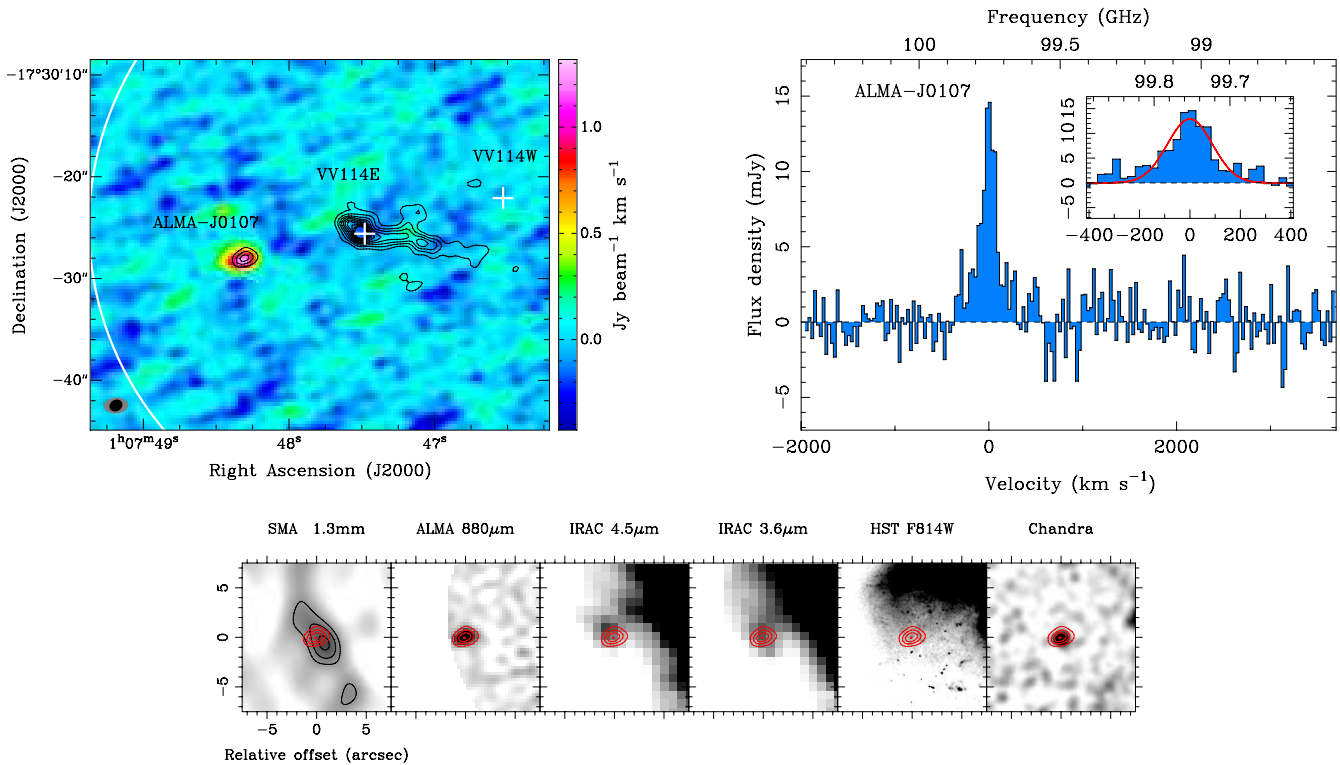}
\caption{
(Top left) Image of ALMA J010748.3$-$173028 in 880-$\micron$ continuum emission (contours) overlaid on 
\textcolor{red}{the} integrated intensity map of 
\textcolor{red}{the} emission line (background).  Images are not corrected for primary-beam attenuation. The contours start at $5\sigma$ with a $5\sigma$ step, where $\sigma = 0.14$ mJy beam$^{-1}$ at the ALMA-J0107 position.  Crosses mark positions of near-infrared peaks shown in Figure~2 of \citet{Imanishi07}.  Partial large circle represents primary beam size at 99.75~GHz.  Filled ellipses in bottom-left corner indicate synthesized beam sizes of emission line image (gray) and 880-$\micron$ images (black).
(Top right) 100 GHz spectrum of ALMA J010748.3$-$173028 across the 1.875~GHz spectral window of ALMA Band~3.  Flux densities are measured with a $4''$ aperture and corrected for primary beam attenuation ($0.79\times$).  Inset shows closeup of the spectrum; solid curve is 
\textcolor{red}{the} best-fit Gaussian. 
(Bottom) $15'' \times 15''$ multiwavelength (millimeter to X-ray) images of ALMA J010748.3$-$173028.  Contours show the ALMA 880-$\micron$ image and start at $4 \sigma$ with a separation of $4 \sigma$, where $\sigma = 0.5$~mJy beam$^{-1}$ is the noise level corrected for primary beam attenuation ($0.28\times$).  Contours of SMA 1.3-mm image are drawn at $2\sigma$, $3\sigma$, and $4\sigma$, where $\sigma = 1.21$~mJy beam$^{-1}$.
}
\label{fig:almaimage}
\end{center}
\end{figure*}


\section{ALMA Observations and Results}

ALMA 3-mm and 880-$\micron$ observations toward VV114 (program ID: 2011.0.00467.S) were conducted using the compact configuration in November 2011 and the extended configuration in May 2012 (3-mm observation only).  The correlator was configured to cover 98.53--102.35 GHz/110.77--113.90 GHz (3~mm) and 323.51--327.25 GHz/335.68--339.31 GHz (880~$\micron$) with a 0.488 MHz resolution.  The 3-mm and 880-$\micron$ primary beam sizes are $62''$ and $19''$, respectively; for the 880-$\micron$ observations, we mosaicked seven pointings to compensate for the small FoV of Band 7. At both 3 mm and 880 $\micron$, Uranus, J1924$-$292, and J0132$-$169 ($6\arcdeg$ away from VV114) were used to calibrate the absolute flux, bandpass, and complex gain, respectively.  
We used \textsc{casa} \citep{McMullin07} to calibrate the visibility data and to image them with a robust weighting of 0.5.  Note that we made the continuum images using only spectral channels that are free from $^{12}$CO emission from VV114, which leaves 6.7-GHz and 6.6-GHz bandwidths at 3 mm and 880 $\micron$, respectively.  We 
\texttt{clean}ed the resulting dirty images down to the 1$\sigma$ level.  The synthesized beam sizes at 3~mm and 880~$\micron$ are $2\farcs 37 \times 1\farcs 57$ (PA = $94\arcdeg$) and $1\farcs 33 \times 1\farcs 12$ (PA = $120\arcdeg$), respectively.   The resulting rms noise levels at 3 mm before correcting for primary beam attenuation were 0.92 mJy~beam$^{-1}$ for a cube with a resolution of 30 km s$^{-1}$ and 50 $\mu$Jy~beam$^{-1}$ for the continuum image.   The 880-$\micron$ noise levels for a 30 km~s$^{-1}$ resolution cube and continuum are 1.5 mJy~beam$^{-1}$ and 0.11 mJy~beam$^{-1}$, respectively.
The flux calibration accuracies in both bands are estimated to be 10\%.

We serendipitously detect a 13$\sigma$ line-emitting object, ALMA-J0107, at 99.753 GHz at $\mathrm{(\alpha,\,\delta)_{J2000} = (01^{h} 07^{m}48\fs 32,\, -17\arcdeg 30' 28\farcs 1)}$, as shown in Figure~\ref{fig:almaimage}.  The image is marginally resolved, and the beam-deconvolved source size measured using a \textsc{casa} task \texttt{imfit} is $1\farcs 6 \pm 0\farcs 2$, although no velocity structure is found.  Figure~\ref{fig:almaimage} (upper right) shows the spectrum of the emission line.  A single Gaussian fit to the spectrum shows $S_\mathrm{peak} = 12.9 \pm 1.0$~mJy, $\Delta v = 210 \pm 18$~km~s$^{-1}$. The integrated intensity is $3.14 \pm 0.15$ Jy~km~s$^{-1}$. 
This line does not correspond to a CO line or any other line of VV114 itself, even if we search a wide velocity range of $\pm 10000$ km s$^{-1}$ around the systemic velocity of VV114 (6100 km~s$^{-1}$).  Moreover, we find no emission line feature in the other spectral windows of Bands 3 and 7 at this position. 
Thus, it is natural to consider a redshifted emission line, especially a $^{12}$CO line, arising from a background galaxy.  We list the possible redshifts, as well as the corresponding CO luminosities and molecular masses, for up to the $J=6$--5 transitions of $^{12}$CO in Table~\ref{table:id}.  The $J=7$--6 ($z=7.087$) transition is ruled out because no [C~\textsc{i}](2--1) at $\nu_\mathrm{obs} = 100.08$~GHz is found. Higher transitions at $z \ge 8.2$ are not plausible.  We will identify the line in \S~\ref{sect:redshiftid}.

The 880-$\micron$ continuum emission in Band~7 is detected at the same position as the Band~3 line peak, close to the edge of the ALMA FoV (Figure~\ref{fig:almaimage}, top left).  The 880-$\micron$ flux density is $11.2 \pm 0.4$ mJy after correcting for primary beam attenuation, whereas we fail to detect the 3-mm continuum in Band 3 down to the 3$\sigma$ upper limit of $< 0.19$ mJy.  The flux density is typical of SMGs \citep{Blain02}, and the inferred far-infrared (FIR) luminosity is $L_\mathrm{FIR} \simeq 1 \times 10^{13} L_\sun$ for $1 < z < 10$ if we assume a dust temperature of $T_\mathrm{dust} = 40$~K and an emissivity index of $\beta = 1.5$.
If the FIR luminosity is powered by starburst activities, a star formation rate is estimated to be $\sim 2\times 10^3 M_\sun$~yr$^{-1}$ following \citet{Kennicutt98}.  We also find 1.3-mm continuum emission in published Submillimeter Array (SMA) data \citep{Wilson08}.  The 1.3-mm flux density is $5.2 \pm 1.3$~mJy. The Rayleigh--Jeans slope is constrained primarily by the ALMA observations, and the lower limit of the spectral index\footnote{The spectral index $\alpha$ is defined such that $S_{\nu} \propto \nu^{\alpha}$.} is $\alpha = 3.5$, consistent with those found in dusty star-forming galaxies.

At the ALMA position, many ancillary data are available from the radio to the X-ray bands.  Table~\ref{table:photometry} lists the results of multiwavelength photometry, and Figure~\ref{fig:almaimage} (bottom) shows multiwavelength images.  Unfortunately, heavy blending by VV114 affects the infrared to optical images, but we clearly see a \emph{Chandra} X-ray counterpart and marginally detect it in the \emph{Spitzer}/IRAC bands.  It is not clear from the current data whether the source is gravitationally lensed.



\begin{deluxetable}{cccccc}
\tablewidth{0.48\textwidth}
\tablecaption{Line Identification\label{table:id}}
\tablehead{
\colhead{$^{12}$CO} & \colhead{$z$} & 
\colhead{$L'_\mathrm{CO}$\tablenotemark{a}} & 
\colhead{$M(\mathrm{H_2})$\tablenotemark{b}} & 
\multicolumn{2}{c}{Predicted $S_\mathrm{CO}\Delta v$\tablenotemark{c}} \\
\colhead{Transition} & & & & 
\colhead{M82\tablenotemark{d}} & \colhead{BR1202\tablenotemark{e}}
}
\startdata
$J = 1\rightarrow 0$ & 0.1556 & 0.37 & 0.29 & 0.6  & 1.1 \\
$J = 2\rightarrow 1$ & 1.311  &  7.1 & 5.7  & 0.8  & 1.6 \\
$J = 3\rightarrow 2$ & 2.467  &  9.9 & 7.9  & 1.1  & 2.4 \\
$J = 4\rightarrow 3$ & 3.622  & 10.5 & 8.4  & 1.4  & 3.5 \\
$J = 5\rightarrow 4$ & 4.777  & 10.4 & 8.3  & 1.7  & 4.9 \\
$J = 6\rightarrow 5$ & 5.932  & 10.0 & 8.0  & 2.0  & 6.6 
\enddata
\tablenotetext{a}{CO line luminosity in units of $10^{10}$~K~km~s$^{-1}$ pc$^2$}
\tablenotetext{b}{Molecular gas mass in units of $10^{10}~M_{\sun}$, derived using a conversion factor of $0.8\,M_\sun$ (K km s$^{-1}$ pc$^2$)$^{-1}$ \citep{Downes98} and assuming thermally excited lines}
\tablenotetext{c}{Integrated intensity in units of Jy km s$^{-1}$, predicted from the 880-$\micron$ flux. See details in \S~\ref{sect:lineid}.}
\tablenotetext{d}{$T_\mathrm{dust} = 40$~K, $\beta = 1.5$, and the CO excitation ladder of M82 \citep{Weiss05} are assumed.}
\tablenotetext{e}{The same as (b), but a dust temperature of 50 K and the CO excitation ladder of BR~1202$-$0725 SE \citep{Salome12} are assumed.}
\end{deluxetable}


\section{Redshift Identification}
\label{sect:redshiftid}

\subsection{Photometric Redshift Estimates}
\label{sect:photoz}

\begin{figure}[htbp]
\begin{center}
\includegraphics[angle=0,width = 0.48\textwidth,keepaspectratio,clip]{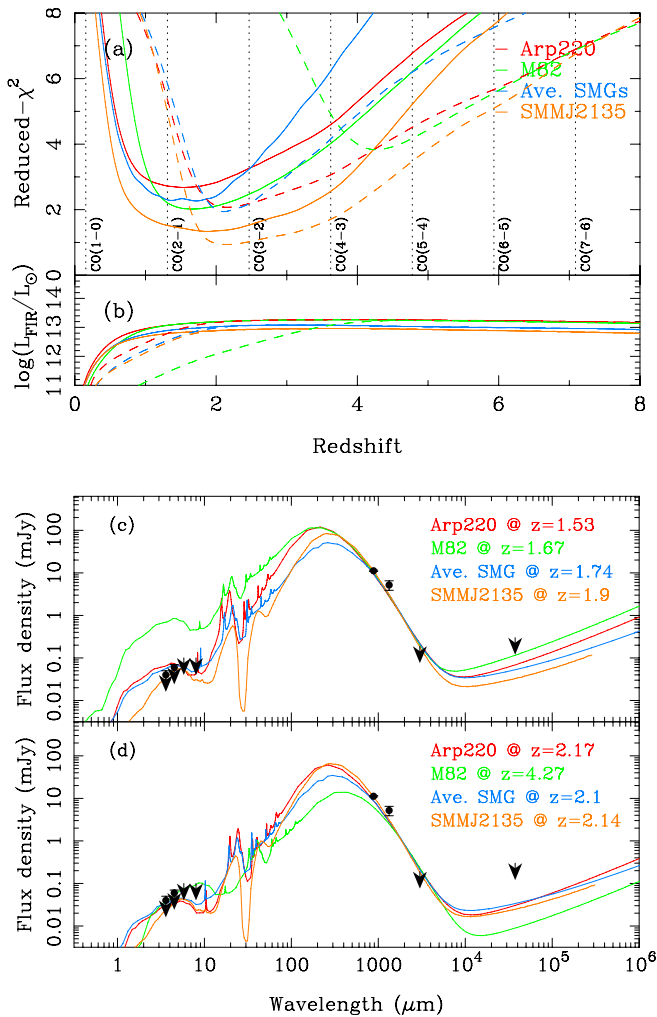}
\caption{Photometric redshift estimates using broadband photometry and templates of spectral energy distributions (SEDs) of Arp~220 (red), M82 (green), composite SMG (blue), and SMM~J2135$-$0201 (orange).  (a) Reduced-$\chi^2$ and (b) far-infrared luminosity $L_\mathrm{FIR}$ as a function of redshift.  Solid curves are obtained using only the photometric data and constraints at $\lambda \ge 880~\micron$; dashed curves are computed using the 3.6- and 4.5-$\micron$ limits as well.  
(c) Broadband photometry and best-fit SED templates obtained using the photometry at $\lambda \ge 880~\micron$.  (d) Same as (c), but the IRAC 3.6 and 4.5-$\micron$ constraints are also used.}
\label{fig:photoz}
\end{center}
\end{figure}

To obtain rough estimates of the redshift, we use four template spectral energy distributions (SEDs) of well-studied starburst galaxies: Arp~220, M82 \citep{Silva98}, SMM~J2135$-$0102 \citep{Swinbank10}, and a composite of radio-identified SMGs \citep{Michalowski10}. Then we fit the submm to radio data to the templates.  Figures~\ref{fig:photoz}a, b, and c show the minimum $\chi^2$ and FIR luminosity as a function of the redshift and the best-fit SEDs, respectively.  The resulting redshifts and the 90\% confidence intervals are $z = 1.53^{+2.31}_{-0.95}$ (Arp 220), $1.67^{+2.12}_{-0.79}$ (M82), $1.90^{+2.25}_{-1.35}$ (SMM~J2135), and $1.74^{+1.28}_{-1.16}$ (mean SMG).  Thus, the likely redshift range is $0.6 < z < 4.1$ for all the templates, suggesting that the $^{12}$CO redshift can be $z = 1.31$, 2.47, or 3.62.  The FIR luminosity, $\log(L_\mathrm{FIR}/L_\sun) \simeq 12.5$--13, is mostly insensitive to the redshift.

The constraint could become tighter when we use the IRAC photometry in addition to the submm to radio data, although the use of IRAC data should be considered cautiously because IRAC observes the rest-frame optical component, which depends strongly on the stellar population model.  The dashed curves in the $\chi^2$ plot of Figure~\ref{fig:photoz}a show the minimum $\chi^2$ values from SED fits using the constraints at 3.6~$\micron$ and 4.5~$\micron$.  The best-fit SEDs are shown in Figure~\ref{fig:photoz}d. The resulting redshifts are $z = 2.17^{+2.31}_{-0.70}$ (Arp 220), $4.27^{+1.96}_{-0.89}$ (M82), $2.14^{+2.37}_{-0.63}$ (SMM~J2135), and $2.10^{+1.46}_{-0.54}$ (mean SMG), although we note that the M82 fit might not be reliable in this case because of poor $\chi^2$ values.  The plausible redshift range in this case is therefore $1.5 < z < 4.5$.
Consequently, the photometric redshift analysis favors a redshift of $z = 2.467$, whereas $z = 3.622$ is within an acceptable range in terms of the $\chi^2$ values.



\begin{deluxetable}{llcc}
\tablewidth{0.48\textwidth}
\tablecaption{Multiwavelength Counterparts to ALMA-J0107 \label{table:photometry}}
\tablehead{
\colhead{Instrument} & \colhead{Band} & \colhead{Flux Density} & \colhead{Unit}
}
\startdata
VLA\tablenotemark{a}&       3.0 cm  & $< 0.3$  (3$\sigma$) & mJy \\
ALMA/Band 3         &       3.0 mm  & $< 0.19$ (3$\sigma$) & mJy \\
SMA                 &       1.3 mm  & $5.2 \pm 1.3$ & mJy \\
ALMA/Band 7         & 880 $\micron$ & $11.2 \pm 0.4$ & mJy \\
\emph{Spitzer}/IRAC & 8.0 $\micron$ & $< 0.1$ & mJy \\
\emph{Spitzer}/IRAC & 5.8 $\micron$ & $< 0.1$ & mJy \\
\emph{Spitzer}/IRAC & 4.5 $\micron$ & $0.06 \pm 0.01$\tablenotemark{b} & mJy \\
\emph{Spitzer}/IRAC & 3.6 $\micron$ & $0.04 \pm 0.01$\tablenotemark{b} & mJy \\
\emph{Chandra}/ACIS & 0.5--10 keV & $2.73 \times 10^{-15}$ & erg s$^{-1}$ cm$^{-2}$ \\
\emph{Chandra}/ACIS & 0.5--2 keV  & $8.34 \times 10^{-16}$ & erg s$^{-1}$ cm$^{-2}$ \\
\emph{Chandra}/ACIS & 2--10 keV   & $1.90 \times 10^{-15}$ & erg s$^{-1}$ cm$^{-2}$ 
\enddata
\tablenotetext{a}{The data were retrieved from the VLA archive.}
\tablenotetext{b}{The flux density is strongly affected by contamination from VV114 and thus should be regarded as an upper limit.}
\end{deluxetable}


\subsection{Line Identification}
\label{sect:lineid}

To confirm that this is a $^{12}$CO line, we estimate the $^{12}$CO intensities using the 880-$\micron$ continuum intensity and some empirical relations and quantities found in a local starburst galaxy, M82. 
From the 880-$\micron$ flux density, we obtain $L_\mathrm{FIR} \sim 9 \times 10^{12} L_\sun$ for $z > 1$ and $\sim 1 \times 10^{12} L_\sun$ for $z = 0.156$ if $T_\mathrm{dust} = 40$~K and $\beta = 1.5$.  The inferred FIR luminosity is almost independent of the redshift at $z > 1$.  Although these are very crude estimates, we use the $L_\mathrm{FIR}$-to-$L'_\mathrm{CO(3-2)}$ correlation \citep{Iono09} to obtain the $^{12}$CO(3--2) luminosity and then the intensities at the possible redshifts.
We find that $S_\mathrm{CO(3-2)} \Delta v \simeq 3.9$, 1.6, 1.1, 1.0, 0.98, and 1.0 Jy km s$^{-1}$ at $z = 0.156$, 1.31, 2.47, 3.62, 4.78, and 5.93, respectively.  Then we assume the CO excitation ladder found in M82 \citep{Weiss05} to obtain the CO intensities at other transitions, which allows us to estimate those at the possible redshifts.  
We repeat this procedure for a higher dust temperature ($T_\mathrm{dust} = 50$~K) and the CO excitation found for BR1202$-$0725 SE \citep{Salome12}, in which the heating of the interstellar medium is dominated by a powerful AGN.
The results are given in Table~\ref{table:id}.
The line intensities are on the order of 1 Jy km s$^{-1}$ and are in good agreement with the observed ones, strongly suggesting that the line is $^{12}$CO because non-$^{12}$CO lines, such as $^{13}$CO and HCN, are $\ge 1$ order(s) of magnitude weaker than $^{12}$CO.  The atomic carbon [CI](1--0) line is possible (at $z = 3.93$) but less likely because the intensity is typically 1/3 to 1/10 that of $^{12}$CO(3--2) \citep[e.g.,][]{Weiss05a}.

Other possible line attributions include the H$_2$O molecule, which is known to have emission lines as bright as those of $^{12}$CO in the submm band.  The major transitions exhibiting strong emission at $\nu_\mathrm{rest} < 1000$~GHz are $J_{K_a,K_c} = 2_{11}$--$2_{02}$ and $2_{02}$--$1_{11}$ at $\nu_\mathrm{rest} = 752.0$ and 987.9 GHz, respectively.  However, the redshifts inferred from the H$_2$O lines would be 6.54 and 8.90, which are outside of the photometric redshift range.  Therefore, these results in combination with the photometric redshift indicate that the line is most likely a redshifted $^{12}$CO transition at $z = 2.467$ or 3.622.



\section{Hard X-ray Detection}

We used the primary package of \emph{Chandra}/ACIS-I data (sequence No.\ 600501, observation ID: 7063).  The X-ray source at the ALMA position also appears $10''$ east of VV114E in Figure 6 of \citet{Grimes06}.  The effective exposure time at the position of ALMA-J0107 was 59~ks.  We count the X-ray events with a $4''$-diameter aperture and measure the background level over a $2'$-diameter circular region centered at ALMA-J0107 while masking VV114 and ALMA-J0107.  We find $\simeq 100$ counts within the aperture, and background subtraction leaves $\simeq 35$ counts, which is poor statistics but just enough to simply model the spectrum.  We use \textsc{xspec} \citep[version 12.8.1,][]{Arnaud96} for spectral modeling and assume a power-law spectrum for the intrinsic spectrum.  For simplicity, we consider only an absorbed spectrum with the intrinsic photon index ($\Gamma$) and obscuring column density ($N_\mathrm{H}$) as free parameters, and eliminate other components such as the scattered spectrum.  Note that this simple assumption may underestimate the obscuring column.

The results for the likely redshifts of $z = 2.47$ and 3.62, as well as for $z = 1.31$ for reference, are shown in Table~\ref{table:xrays}.  The inferred unabsorbed luminosity, $L_\mathrm{X}$, covers the range $43.5 \le \log(L_\mathrm{X}/\mathrm{erg~s^{-1}}) \le 44.6$ for $1.31 \le z \le 3.62$, which is comparable to that of $z \sim 2$ AGN-classified SMGs \citep{Alexander05}.  Using the method described in \citet{Tamura10}, we estimate an AGN bolometric luminosity of $L_\mathrm{bol} \sim (0.8$--$34) \times 10^{11} L_\sun$ for $1.31 \le z \le 3.62$.  Despite a large uncertainty, this gives a mass of $\sim (0.1$--$5) \times 10^{8} (\eta_\mathrm{Edd}/0.2)^{-1} M_{\sun}$ for the accreting supermassive black hole, where $\eta_\mathrm{Edd}$ is the Eddington ratio, which is typically 0.2--0.6 in SMGs \citep{Alexander08}. The bolometric luminosity is comparable to those found in the most luminous AGNs in the local Universe (e.g., Mrk~231), but it is even less than the FIR luminosity of the host galaxy ALMA-J0107 ($L_\mathrm{FIR} \simeq 9 \times 10^{12} L_\sun$ for $T_\mathrm{dust} = 40$~K), suggesting that the large FIR luminosity of ALMA-J0107 is not dominated by the AGN but can be attributed to massive star formation activities.



\begin{deluxetable}{cccccc}
\tablewidth{0.48\textwidth}
\tablecaption{X-ray Properties of ALMA-J0107 at Plausible Redshifts \label{table:xrays}}
\tablehead{
\colhead{$z$} & \colhead{$\log{N_\mathrm{H}}$} & \colhead{$\Gamma$} & \colhead{$E$\tablenotemark{a}} & \colhead{$L_\mathrm{X}$\tablenotemark{b}} & \colhead{$L_\mathrm{bol}$} \\
& \colhead{(cm$^{-2}$)} &  & (keV) & \colhead{(ergs~s$^{-1}$)} & \colhead{($10^{11}L_\sun$)} 
}
\startdata
1.31 & $22.8^{+0.5}_{-1.3}$   & $2.6^{+3.0}_{-1.4}$ & 1.2--23 & $3.0 \times 10^{43}$ & 0.8 \\
2.47 & $23.2^{+0.6}_{-1.5}$   & $2.5^{+2.6}_{-1.3}$ & 1.7--35 & $1.4 \times 10^{44}$ & 7.4 \\
3.62 & $23.6^{+0.8}_{-23.6}$  & $2.4^{+3.4}_{-1.3}$ & 2.3--46 & $3.9 \times 10^{44}$ & 34 
\enddata
\tablenotetext{a}{Rest-frame energy band corresponding to the observed-frame 0.5--10 keV band.}
\tablenotetext{b}{Unabsorbed luminosity in energy band $E$.}
\end{deluxetable}


\section{Number Counts of CO Emitters}

How frequently is a $\simeq$3 Jy km s$^{-1}$ $^{12}$CO emitter observed by chance?  We use a mock galaxy catalog from the \textsc{s$^3$~sax} simulation \citep{Obreschkow09a,Obreschkow09b} to estimate the expected number of detections of redshifted $^{12}$CO, regardless of redshift, with a single pointing/tuning of ALMA at 100 GHz. This is a semi-analytic simulation of neutral atomic (H~\textsc{i}) and molecular (H$_2$) hydrogen in galaxies and the associated CO lines; it is based on the Millennium Simulation \citep{Springel05}, which reliably recovers galaxies with cold hydrogen masses $M$(\textsc{H~i}+H$_2) > 10^8 M_\sun$.  It reproduces the local CO(1--0) luminosity function well \citep{Keres03}, whereas those at high-$z$ are not fully verified by observations;
It may underpredict the number density of $\sim 0.6$ Jy km s$^{-1}$ CO emitters at $z \sim 1.5$ by a factor of several, in comparison with CO observations of $z \sim 1.5$ BzK galaxies \citep{Daddi08, Daddi10}.  The cumulative number counts of CO emitters, $\mathcal{N}(>\!S\Delta v)$, expected in a bandwidth $d\nu_\mathrm{obs}$ and a primary beam solid angle $d\Omega$ are described as 
\begin{eqnarray}
\mathcal{N}(>\!S\Delta v) &=& \sum_{J=1}^{\infty} \left( \frac{dN}{dz} \right)_{z_J} dz\,d\Omega \\
&=& \frac{d\nu_\mathrm{obs}\,d\Omega}{\nu_\mathrm{obs}} 
    \sum_{J=1}^{\infty} \left( \frac{dN}{dz} \right)_{z_J} (1+z_J),
\end{eqnarray}
where $z_J$ is the redshift at which the $J\rightarrow J-1$ transition of CO is observed at $\nu_\mathrm{obs}$, and $(dN/dz)_{z_J}$ is the surface number density of galaxies observed in the $J\rightarrow J-1$ transition per redshift interval with line fluxes above a certain threshold, $S\Delta v$.  To estimate $dN/dz$ at each $z_J$, we extract sources with an integrated intensity higher than 1 Jy km s$^{-1}$ from the simulated volumes defined by an area of $62.5 \times 62.5\,h^{-2}$ Mpc$^2$ with a depth of $\Delta z = 0.50$ at $z = 1.31$, 2.47, 3.62, 4.78, and 5.93 (we choose $\Delta z = 0.10$ for $z = 0.156$).

Consequently, we expect $\simeq 0.011$ source with $>1$ Jy~km~s$^{-1}$ per ALMA FoV (2800 arcsec$^2$) and bandwidth (7.5 GHz).  Most of the sources ($\simeq 90$\%) are CO(1--0) at $z = 0.156$.  The remaining 10\% are almost evenly distributed at $z = 1.31$, 2.47, and 3.62, but no source is found at $z \ge 4.78$.  The SED analysis (\S~\ref{sect:photoz}) rules out the lowest redshift ($z = 0.156$), even though the probability of a chance detection appears to be highest for $z = 0.156$.  Although it should be properly tested whether the \textsc{s$^3$~sax} simulation reproduces the brightest ($> 1$ Jy~km~s$^{-1}$) population of CO emitters at $z > 1$, this result implies that ALMA-J0107, likely at $z = 2.47$ or 3.62, is a very rare galaxy that falls within the ALMA bandwidth by chance (one out of $\sim 1000$ FoVs).


\section{Summary}

We presented the detection of a $^{12}$CO-emitting galaxy, ALMA-J0107, beyond the nearby merging galaxies VV114.  The integrated intensity of CO and the 880-$\micron$ flux density are $3.14 \pm 0.15$~Jy km s$^{-1}$ and $11.2 \pm 0.5$ mJy, respectively.  The photometric redshift analysis favors $z = 2.467$, but $z = 3.622$ is acceptable.  The molecular mass and FIR luminosity at the plausible redshifts are $M(\mathrm{H_2}) \sim 8 \times 10^{10} M_{\sun}$ and $L_\mathrm{FIR} \sim 1 \times 10^{13} L_{\sun}$, respectively, which correspond to a star formation rate of $\sim 2000 M_{\sun}$ yr$^{-1}$.
We identified a hard X-ray source at the ALMA position, suggesting the presence of a luminous ($L_\mathrm{X} \sim 10^{44}$ erg s$^{-1}$) AGN behind a large hydrogen column ($23.2 \lesssim \log{[N_\mathrm{H}/\mathrm{cm^{-2}}]} \lesssim 23.6$ for the likely redshifts). 
However, the intrinsic properties of the AGN (e.g., the bolometric luminosity) depend strongly on the redshift, although the FIR luminosity and molecular mass are rather insensitive to the redshift.  This fact makes it difficult to investigate the power source of ALMA-J0107 and the evolutionary status of black hole growth in ALMA-J0107. It is obviously quite important to confirm the redshift through observations of the other transitions of $^{12}$CO.

This serendipitous detection of a CO-emitting galaxy demonstrates that ALMA is capable of identifying an emission-line galaxy such as ALMA-J0107.  We have shown that the likelihood of stumbling across such a source is not high, and redshift determination remains a challenge even when one line and the continuum are clearly detected.  Nevertheless, $\sim 1000$ pointings of ALMA Band 3 will offer an additional detection of a $> 1$~Jy km s$^{-1}$ CO source at high redshift ($z > 1$).  A CO emitter at this flux level can routinely be detected at 100~GHz in only a few minutes with the full ALMA if the line happens to fall in the observing band.  A complete census of background high-$z$ CO emitters in Band 3 archival cubes, as well as \textsc{[C~ii]} emitters in Band 6/7, is encouraged.


\acknowledgments
We acknowledge the anonymous referee for useful comments.  We thank C.~D.\ Wilson for providing the SMA image.  Y.T.\ thanks T.\ Kawaguchi for fruitful discussions.
This work was supported by JSPS KAKENHI Grant Number 25103503.  
This paper makes use of the following ALMA data: ADS/JAO.ALMA\#2011.0.00467.S. ALMA is a partnership of ESO (representing its member states), NSF (USA), and NINS (Japan), together with NRC (Canada) and NSC and ASIAA (Taiwan), in cooperation with the Republic of Chile.  The Joint ALMA Observatory is operated by ESO, AUI/NRAO, and NAOJ. NRAO is a facility of the NSF operated under cooperative agreement by Associated Universities, Inc.

{\it Facilities:} \facility{VLA}, \facility{ALMA}, \facility{SMA}, \facility{\emph{Spitzer} (IRAC)}, \facility{CXO (ASIS)}.





\begin{thebibliography}{}
\bibitem[Alexander et al.(2005)]{Alexander05} Alexander, D.~M., Bauer, F.~E., Chapman, S.~C., et al.\ 2005, \apj, 632, 736  
\bibitem[Alexander et al.(2008)]{Alexander08} Alexander, D.~M., Brandt, W.~N., Smail, I., et al.\ 2008, \aj, 135, 1968 
\bibitem[Arnaud(1996)]{Arnaud96} Arnaud, K.~A.\ 1996, Astronomical Data Analysis Software and Systems V, 101, 17 
\bibitem[Blain et al.(2002)]{Blain02} Blain, A.~W., Smail, I., Ivison, R.~J., Kneib, J.-P., \& Frayer, D.~T.\ 2002, \physrep, 369, 111 
\bibitem[Daddi et al.(2008)]{Daddi08} Daddi, E., Dannerbauer, H., Elbaz, D., et al.\ 2008, \apjl, 673, L21 
\bibitem[Daddi et al.(2010)]{Daddi10} Daddi, E., Bournaud, F., Walter, F., et al.\ 2010, \apj, 713, 686 
\bibitem[Downes \& Solomon(1998)]{Downes98} Downes, D., \& Solomon, P.~M.\ 1998, \apj, 507, 615 
\bibitem[Grimes et al.(2006)]{Grimes06} Grimes, J.~P., Heckman, T., Hoopes, C., et al.\ 2006, \apj, 648, 310 
\bibitem[Hatsukade et al.(2013)]{Hatsukade13} Hatsukade, B., Ohta, K., Seko, A., Yabe, K., \& Akiyama, M.\ 2013, \apjl, 769, L27 
\bibitem[Imanishi et al.(2007)]{Imanishi07} Imanishi, M., Nakanishi, K., Tamura, Y., Oi, N., \& Kohno, K.\ 2007, \aj, 134, 2366 
\bibitem[Iono et al.(2013)]{Iono13} Iono, D., Saito, T., Yun, M.~S., et al.\ 2013, \pasj, 65, L7 
\bibitem[Iono et al.(2009)]{Iono09} Iono, D., Wilson, C.~D., Yun, M.~S., et al.\ 2009, \apj, 695, 1537 
\bibitem[Kennicutt(1998)]{Kennicutt98} Kennicutt, R.~C., Jr.\ 1998, \araa, 36, 189 
\bibitem[Keres et al.(2003)]{Keres03} Keres, D., Yun, M.~S., \& Young, J.~S.\ 2003, \apj, 582, 659 
\bibitem[McMullin et al.(2007)]{McMullin07} McMullin, J.~P., Waters, B., Schiebel, D., Young, W., \& Golap, K.\ 2007, Astronomical Data Analysis Software and Systems XVI, 376, 127 
\bibitem[Micha{\l}owski et al.(2010)]{Michalowski10} Micha{\l}owski, M., Hjorth, J., \& Watson, D.\ 2010, \aap, 514, A67 
\bibitem[Obreschkow et al.(2009a)]{Obreschkow09a} Obreschkow, D., Heywood, I., Kl{\"o}ckner, H.-R., \& Rawlings, S.\ 2009a, \apj, 702, 1321 
\bibitem[Obreschkow et al.(2009b)]{Obreschkow09b} Obreschkow, D., Kl{\"o}ckner, H.-R., Heywood, I., Levrier, F., \& Rawlings, S.\ 2009b, \apj, 703, 1890 
\bibitem[Salom{\'e} et al.(2012)]{Salome12} Salom{\'e}, P., Gu{\'e}lin, M., Downes, D., et al.\ 2012, \aap, 545, A57 
\bibitem[Silva et al.(1998)]{Silva98} Silva, L., Granato, G.~L., Bressan, A., \& Danese, L.\ 1998, \apj, 509, 103 
\bibitem[Springel et al.(2005)]{Springel05} Springel, V., White, S.~D.~M., Jenkins, A., et al.\ 2005, \nat, 435, 629 
\bibitem[Swinbank et al.(2010)]{Swinbank10} Swinbank, A.~M., Smail, I., Longmore, S., et al.\ 2010, \nat, 464, 733 
\bibitem[Swinbank et al.(2012)]{Swinbank12} Swinbank, A.~M., Karim, A., Smail, I., et al.\ 2012, \mnras, 427, 1066
\bibitem[Tamura et al.(2010)]{Tamura10} Tamura, Y., Iono, D., Wilner, D.~J., et al.\ 2010, \apj, 724, 1270 
\bibitem[Vieira et al.(2013)]{Vieira13} Vieira, J.~D., Marrone, D.~P., Chapman, S.~C., et al.\ 2013, \nat, 495, 344 
\bibitem[Wagg et al.(2012)]{Wagg12} Wagg, J., Wiklind, T., Carilli, C.~L., et al.\ 2012, \apjl, 752, L30
\bibitem[Wang et al.(2013)]{Wang13} Wang, R., Wagg, J., Carilli, C.~L., et al.\ 2013, \apj, 773, 44 
\bibitem[Wei{\ss} et al.(2005a)]{Weiss05a} Wei{\ss}, A., Downes, D., Henkel, C., \& Walter, F.\ 2005a, \aap, 429, L25 
\bibitem[Wei{\ss} et al.(2005b)]{Weiss05} Wei{\ss}, A., Walter, F., \& Scoville, N.~Z.\ 2005b, \aap, 438, 533 
\bibitem[Wilson et al.(2008)]{Wilson08} Wilson, C.~D., Petitpas, G.~R., Iono, D., et al.\ 2008, \apjs, 178, 189 
\end{thebibliography}
\end{document}